\def\simlt{$\; \buildrel < \over \sim \;$}
\def\ltsima{\lower.5ex\hbox{\simlt}}
\def\simgt{$\; \buildrel > \over \sim \;$}

\def\ltsima{$\; \buildrel < \over \sim \;$}
\def\simlt{\lower.5ex\hbox{\ltsima}}
\def\gtsima{$\; \buildrel > \over \sim \;$}
\def\simgt{\lower.5ex\hbox{\gtsima}}

\def\etal{{\it et~al.~}}

\def\exosat{{\it EXOSAT~}}
\def\asca{{\it ASCA~}}
\def\ginga{{\it Ginga~}}
\def\rosat{{\it ROSAT~}}
\def\xmm{{\it XMM-Newton~}}
\def\sax{{\it BeppoSAX~}}
\def\chandra{{\it Chandra~}}
\def\rxte{{\it RXTE~}}
\def\asca{{\it ASCA~}}

\def\flux{\;\rm erg\ cm^{-2}\ s^{-1}}
\def\csq{$\chi^2$}

\def\flux{\rm erg\ cm^{-2}\ s^{-1}}
\def\509{Mkn~509}
 
\documentclass{aa}
\usepackage{graphicx}

\begin{document}

%\thesaurus{ }     

\title{Close and distant reprocessing media in \509 studied with \sax}

\author{ A. De Rosa \inst{1}
\and L. Piro \inst{1}
\and G. Matt \inst{2}
\and G.C. Perola \inst{2}
}
\offprints{Alessandra De Rosa: derosa@rm.iasf.cnr.it}     

\institute{
{Istituto di Astrofisica Spaziale e Fisica Cosmica,
C.N.R.,Via Fosso del Cavaliere, Roma, Italy}
\and
{Dipartimento di Fisica, Universit\`a degli Studi ``Roma Tre'', Via della Vasca Navale 84, I--00146 Roma, Italy}
}
\date{Received ; Accepted }

\abstract{
We present the broad band analysis of two \sax observations of the 
Seyfert 1 \509. In 2000 the source was in a typical 
flux state, F$_{2-10keV}=2.7\times 10^{-11} \flux$, while in 1998 it was 
found in a high flux state, F$_{2-10keV}=5.7\times 10^{-11} \flux$.   
A comparison between the two states shows a energy--dependent flux variation: 
of about a factor three and a factor two in the LECS (0.15-3 keV) and MECS (1.5-10 keV), 
respectively, while in the PDS (13-200 keV) the difference is marginal.
A soft excess, a narrow iron line and a Compton reflection 
hump above 10 keV, are clearly apparent in the residuals after fitting the spectra
with a simple power law.
We tested two alternative models. In the first the iron line and 
the high energy 
bump are well reproduced by reprocessing in a cold and Compton thick
material. The intensity of the iron line (also consistent with a \chandra
measurement) as well as the normalization of the reflection hump are consistent
with a constant in the two epochs: this, combined with the fact that 
the line is narrow as observed by \chandra, suggests a common
origin from distant and optically thick matter. This model further
requires a component to model the soft excess: the empirical choice of two
black bodies accounts well for the excess in both observations, their
combined strength was a factor about three higher in the high than in
the low flux state defined above.
However the relative contribution of the soft excess 
is higher in the low flux state.
In the second model we attempted to reproduce all spectral features,
except for the narrow cold line, with reflection from an ionized
disc. This model is successful only in the high flux state,
but it fails in the low flux state, when the soft excess is 
only partially accounted for.
In either models, the slope of the power law is larger in the
high than in the low flux state, ($\Delta\Gamma \sim$0.2), in
agreement with a behaviour known to be shared by several objects 
of the same type.

} 

\titlerunning{Close and distant reprocessing media in \509}
\maketitle

\section{Introduction}

The most popular model for the X-ray emission in AGN envisages a 
hot, optically-thin plasma (the corona) which up-scatters to X-ray energies 
the soft photons produced in an optically thick accretion disc 
(e.g. Haardt \& Maraschi 1993).
The disc, in its turn, reprocesses and re--emits part of the
Comptonized flux, so producing the characteristic reprocessing features, 
the Compton reflection hump and the fluorescent iron K$\alpha$ line.
The presence of a broad component of the iron line (Fabian \etal 2000) in the 
X-ray spectra of Seyfert galaxies was first observed       
by \asca (Tanaka \etal 1995, Nandra \etal 1997, Yaqoob \etal 2002), 
confirmed by \sax (Guainazzi \etal 1999) and by 
\xmm (Fabian \etal 2002, Wilms \etal 2001) observations. However,
\chandra and \xmm observations clearly indicate that a narrow 
component, likely originating in distant matter, 
is also almost always present 
(Reeves \etal 2001, Pounds \etal 2001, Matt \etal 2001, Yaqoob \etal 2001, 
Kaspi \etal 2001, Pounds \etal 2002). 
The BLR and the optically thick torus are the best candidate regions to
produce this narrow component.
In the latter, the matter is likely to be Compton--thick
(N$_H > 10^{24}$ cm$^{-2}$) and a Compton reflection component
is also expected (Ghisellini \etal 1994, Matt et al. 2003), 
as already observed in NGC~4051 (Guanazzi \etal 1998) and 
NGC~5506 (Matt \etal 2001).

In this paper we present the 2000 \sax observations of the Seyfert 1 galaxy
\509, and compare it with the 1998 observation, already 
discussed by Perola \etal (2000).
This source has a typical X-ray luminosity 
(in the 2--10 keV energy range) of 3$\times 10^{44}$ erg s$^{-1}$ 
($H_0$=50 km s$^{-1}$ Mpc$^{-1}$) at 
$z$=0.034, with a Galactic absorption column of
N$_H^{Gal}=4.4\times 10^{20}$ cm$^{-2}$  (Murphy \etal 1996).
A soft excess in the X-ray spectra was observed by a \rosat and \ginga 
simultaneous observation (Pounds \etal 1994), confirming a previous 
\exosat finding (Morini \etal 1987). The 
\asca observation, however, suggested a warm absorber component in 
\509 rather than a true soft excess (Reynolds 1997, George \etal
1998), but \sax (Perola \etal 2000) and \xmm (Pounds \etal 2001) 
confirmed the soft X-ray excess below 1 keV, and
showed also the presence of a narrow iron line
(already detected by \ginga and \asca) and a Compton reflection bump above 
10 keV.

The observations and data reduction are described in Section 2. 
A model--independent variability study is presented in Section 3. 
The spectral analysis is reported in Section 4 and discussed in Section 5.
Our conclusions are drawn in Section 6.

\section{Observations and data reduction}

\509 was observed by \sax (Boella \etal 1997) two times in 
1998 (May 18, October 11)
and four times in 2000 (November 3-8-18-24).
The LECS, MECS and PDS data reduction followed the standard procedure
(Fiore, Guainazzi \& Grandi 1999).
The PDS spectra were filtered with variable rise time.
We extracted the spectrum within circular regions around the 
source centroid with radii of 4' and 8' for the MECS and LECS, respectively.
The background was extracted from event files of source-free regions 
(``blank fields''). Spectral models 
were fitted to the data using the XSPEC package V11.2. All quoted 
uncertainties correspond to 90\% confidence level for one interesting 
parameter ($\Delta$\csq=2.71).
The models included a normalization factor for each instrument
to take into account intercalibrations. 
We fixed the PDS/MECS normalization to 0.8
while the LECS/MECS normalization ratio was
running between 0.7 and 1 (Fiore, Guainazzi \& Grandi 1999).

In Figure \ref{lightcurve} we show the LECS (0.7-2.5 keV), 
MECS (2-10 keV) and PDS (13-200 keV) light curves (the binsize is 6000 s), 
and in the last panel the hardness ratio MECS (2-10 keV)/LECS (0.7-2.5 keV).
The source does not show significant spectral
variations within the two sets of observations, thence
we combined the
two in 1998  as well as the four in
2000. The journal of the combined 
observations is shown in Table \ref{journal}. 
In 1998 \509 was in a relatively bright 
flux state, F$_{2-10 keV}=5.66\times 10^{-11} \flux$,
while in 2000 it was at a more typical
flux level, F$_{2-10 keV}=2.7\times 10^{-11} \flux$.

\begin{table*}
\caption{\509 by \sax. Journal of the observations in 1998 (sum of two) and 
2000 (sum of four).} 
\begin{flushleft}
\begin{tabular}{l c c c c}
\noalign{\hrule}
\noalign{\medskip}
& & & &  \\
OBS & LECS (0.15- 3 keV) & MECS (1.5 -10 keV) & PDS (13- 200 keV) & t$_{exp}$ (MECS) \\
&(s$^{-1}$) &(s$^{-1}$) &(s$^{-1}$) & (s) \\
\hline
& & & &  \\
1998 & 0.464 $\pm$ 0.003 & 0.724 $\pm$ 0.003 & 0.77 $\pm$ 0.03 & 87834 \\
& & & &  \\
2000 & 0.167 $\pm$ 0.002 & 0.325 $\pm$ 0.001 & 0.58 $\pm$ 0.03 & 152777\\
& & & &  \\
\hline
\end{tabular}
\end{flushleft}
\label{journal}
\end{table*}

\section{Spectral ratio}

\begin{figure*}
\centering
\includegraphics[height=6.cm,width=6.cm]{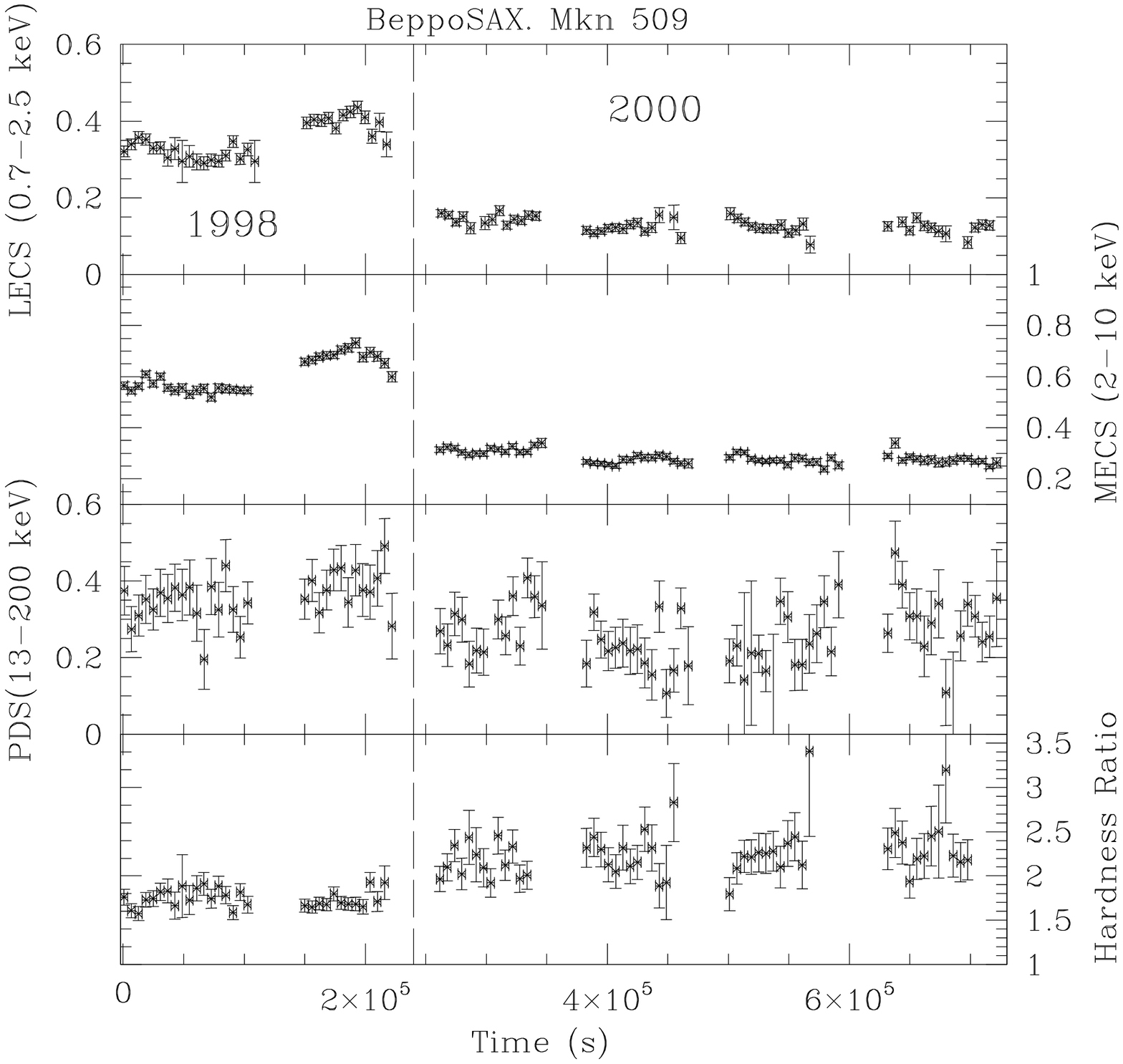}
\hspace{1cm}
\includegraphics[height=6.cm,width=6.cm]{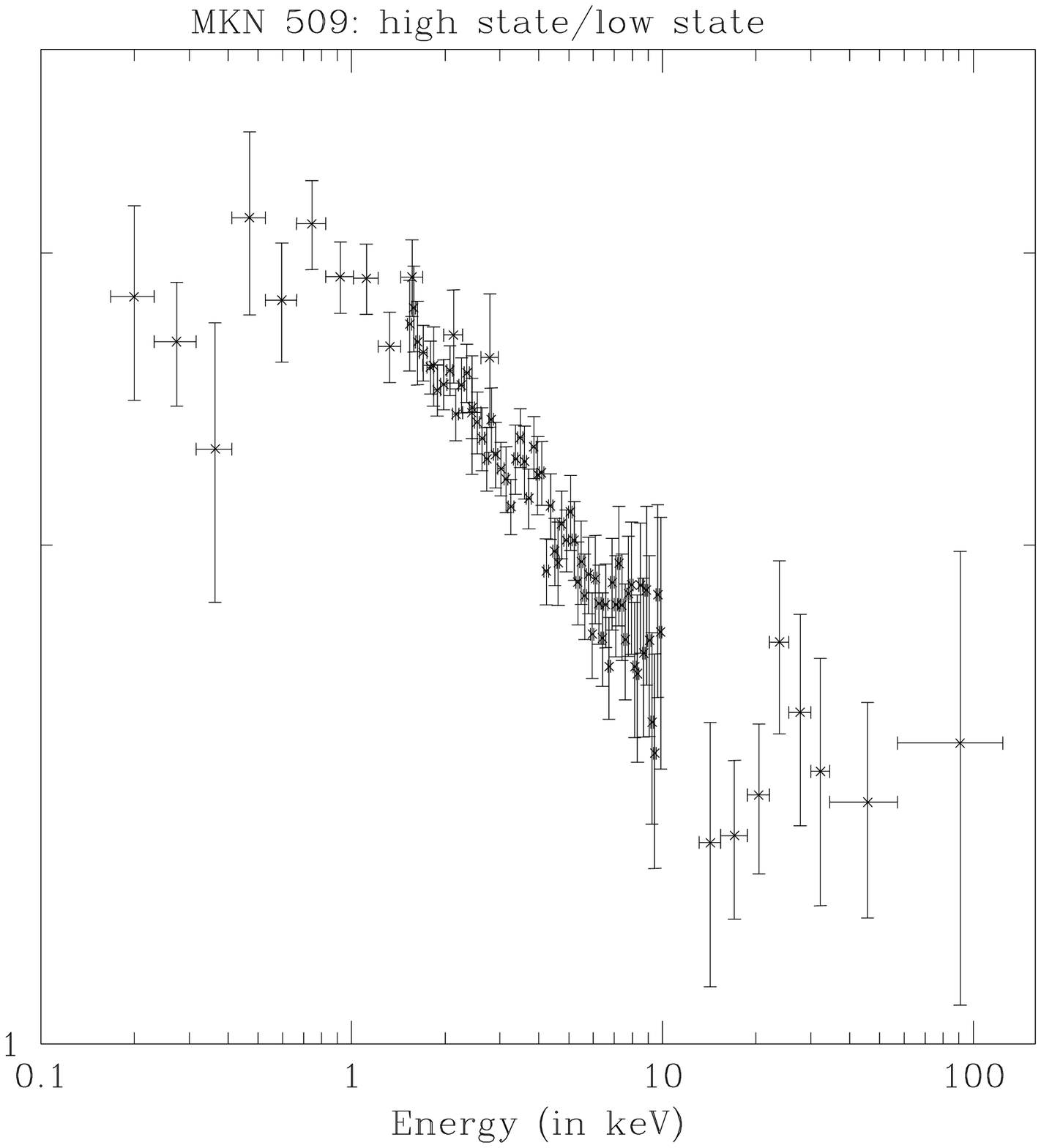}
\caption[]{Left panel. Lightcurves and hardness ratio 
MECS(2-10 keV)/LECS(0.7-2.5 keV) of the \sax observations 
(the binsize is 6000 s). 
The time intervals between the different looks in 1998 and 2000 are fake.
Right panel. Spectral ratio between the high (1998) and low (2000) flux state}
\label{lightcurve}
\end{figure*}

The two states are characterized by different spectral
shapes, as immediately revealed, in a model--independent way,
by the ratio of the spectra (right panel of Figure \ref{lightcurve}):

1) In the 1-10 keV energy range a change in the slope
is apparent (with a pivot point around 70-80 keV), 
the steeper spectrum corresponding to the
larger flux state. Since in this band the incidence of absorption, 
reflection component and soft excess is negligible,
the change must be associated with the primary, power law
component.

2) Below 1 keV the ratio suggests the presence of another component,
which varied in a different way with respect to the power law.

3) In the 10-100 keV range the ratio is consistent with a constant
value. Being this one the region where
Compton reflection is important, this result suggests an approximately
constant intensity in this additional component, despite the change of the
primary flux.

In the next section we will test 
these results with a detailed spectral analysis.

%%%%%%%%%%%%%%%%%%%%% 3-100 keV fit table %%%%%%%%%%%%%%%%%%%%%%%%
%
\begin{table*}
\caption{Best fit spectra between 3-100 keV in the low (2000) and high (1998) 
flux state of \509.}
\begin{flushleft}
\begin{tabular}{lccccccc}
\noalign{\hrule}
\noalign{\medskip}
 & $\Gamma$ & E$_c$(keV) & R & $^\circ$I$_{Fe}$ & $EW_{Fe}$(eV) & E$^{obs}_{Fe}$ (keV) 
& $\chi^2$/dof \\
 & & & & & & &\\
\hline
 & & & & & & &\\
2000 & 1.59$^{+0.07}_{-0.06}$ & 83$^{+47}_{-20}$ & 0.86$^{+0.40}_{-0.31}$ & 
2.8$^{+1.8}_{-1.0}$ & 96$\pm$60 & 6.33$^{+0.19}_{-0.15}$ & 60/66 \\
  & & & & & & & \\
\hline
 & & & & & &  &\\
1998 &1.80$^{+0.05}_{-0.06}$ & 110$^{+40}_{-70}$ & 0.64$^{+0.33}_{-0.26}$ 
& 3.3$^{+1.6}_{-1.7}$ &57$\pm$27 & 6.37$^{+0.19}_{-0.17}$ & 54/66 \\
 & & & & & &  &\\
\hline
\end{tabular}
\end{flushleft}
\small{$^\star$ In 10$^{-5}$ ph cm$^{-2}$ s$^{-1}$ at the line energy.}
\label{fit 3-10 table}
\end{table*}
\begin{figure*}
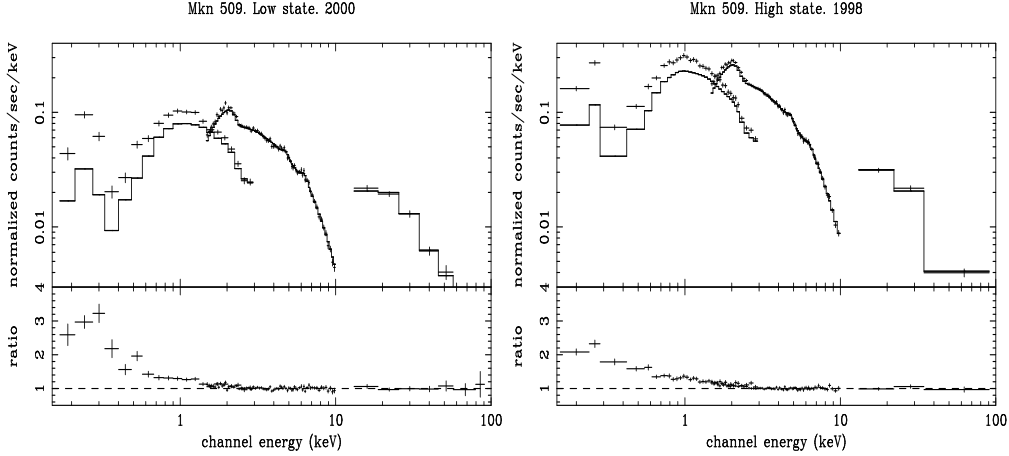

\centering
\includegraphics[height=6.6cm,width=6.cm,angle=-90]{se_2000.ps}
\includegraphics[height=6.6cm,width=6.cm,angle=-90]{se_1998.ps}
\caption[]{Hard X-ray spectral fit for E$>$ 3 keV extrapolated to 
0.15 keV in the \sax-LECS energy band. A soft X-ray excess is present. 
The ratios data/model are showing as this spectral component seems have a
larger contribute in the low flux state of \509 
(left panel, see discussion on the text for details.)} 
\label{soft excess}
\end{figure*}
%
%%%%%%%%%%%%%%%%%% COLD DISC TABLE %%%%%%%%%%%%%%%%%%%%%%
%
%
\begin{table*}
\caption{Cold disc reflection model. Broad band best fit spectra.}
\begin{flushleft}
\begin{tabular}{lccccccccc}
\noalign{\hrule}
\noalign{\medskip}
 & $\Gamma$ & E$_c$(keV) & R & kT$_{BB}^1$ (eV) & kT$_{BB}^2$ (eV) & $^\star$I$_{Fe}$ & $EW_{Fe}$(eV) & $\chi^2$/dof & Null hyp prob \\
 & & & & & & & & &\\
\hline
 & & & & & & & & &\\
2000 & 1.60$\pm^{0.07}_{0.06}$ & 84$^{+36}_{-19}$ & 0.90$\pm^{0.50}_{0.35}$ 
& 71$^{+6}_{-7}$ & 243$^{+27}_{-47}$ & 2.8$\pm$1.1 & 94$\pm$37 & 157.4/138 & 0.124\\
 & & & & & & & & & \\
\hline
 & & & & & & & & &\\
1998 &1.82$\pm$0.06 & 115$^{+92}_{-43}$ & 0.70$^{+0.45}_{-0.25}$ &71$^{+6}_{-8}$ & 260$\pm$25 & 3.4$\pm$2.0 & 62$\pm$36& 127.8/138 & 0.721\\
 & & & & & & & & &\\
\hline
\end{tabular}
\end{flushleft}
\small{$^\star$ In 10$^{-5}$ ph cm$^{-2}$ s$^{-1}$ at the line energy.}
\label{cold disc model}
\end{table*}
%%%%%%%%%%%%%%%%%%%%%%%%%%%%%%%%%%%%%%%%%%%%%%%%%%%%%%%%%
%%%%%%%%%%%%%%%%%%%%%%%%%%%%%%%%%%%%%%%%%%%%%%%%%%%%%%%%%%%%%
\begin{table*}
\caption{The ionized disc reflection model. Broad band best fit spectra. The disc radial 
emissivity law is $\epsilon(r)\propto r^\beta$, with $\beta$=-3. 
The inner and outer radii were kept fixed respectively to r$_{in}$=6 r$_g$ and 
r$_{out}$=10$^3$ r$_g$. with r$_g$=Gm/c$^2$ 
the gravitational radius.}
\begin{flushleft}
\begin{tabular}{l c c c c c c c c c }
\noalign{\hrule}
\noalign{\medskip}
 & \bf $\Gamma$ & log ($\xi$) & $R$ & $^\star I_{Fe}^{narrow}$ & $EW_{Fe}^{narrow}$(eV) &$\chi^2$/dof & Null hyp prob \\
& & & & & & & \\
\hline
& & & & & & &\\
2000 & 1.70$\pm$0.02 & 1.55$^{+0.12}_{-0.04}$ & 1.14$\pm0.15$ & 1.6$\pm$0.9 & 62$\pm$35 & 183.2/143 & 0.013 \\ 
& & & & & & &\\
\hline
& & & & & & &\\
1998 & 1.89$\pm$0.02 & 1.55$^{+0.11}_{-0.22}$ & 0.75$^{+0.14}_{-0.10}$ & 2.3$\pm 1.6$ & 45$\pm$30 & 155.3/143 & 0.228 \\
& & & & & & &\\
\hline
\end{tabular}
\end{flushleft}
\small{$^\star$ In 10$^{-5}$ ph cm$^{-2}$ s$^{-1}$ at the line energy.}
\label{iondisc table}
\end{table*}
%%%%%%%%%%%%%%%%%%%%%%%%%%%%%%%%%%%%%

\section{Broad band spectral analysis}

\subsection{Cold disc reflection model}
\label{cold disc section}

\509 is known to possess a  soft excess, as 
observed by \exosat (Turner \etal 1991), \rosat (Pounds \etal 1994), 
\sax (Perola \etal 2000) and \xmm (Pounds \etal 2001). We first
excluded the energy range where the excess is present, by
fitting the 1998 and 2000 spectra from 3 to 100 keV
with a power law multiplied by exp(--E/E$_c$), a cold reflection 
component (PEXRAV,
Magdziarz \& Zdziarski 1995,  with inclination angle fixed to $i$=30$^\circ$) 
and a Gaussian component to model the iron line.
With the intrinsic width of the line as a free parameter,
the fit yields an unacceptably large value of $\sigma \sim 2$ keV 
in both spectra, that was likely due to a bad description of the 
continuum at the line energy. This parameter is however basically
unconstrained and in fact, with a width fixed to 0.1 keV, the fit is equally
good. We therefore adopted this valuel; the results are
reported for both epochs in Table \ref{fit 3-10 table}.

The $\chi^2$ is fully acceptable in both observations 
($(\chi^2/dof)_{2000}$=60/66, $(\chi^2/dof)_{1998}$=54/66). 
The narrow line energy is consistent in both spectra with neutral iron,
with an intensity that does not appear to vary significantly between the 
low and the high flux states.

A strong soft excess at E$<$ 3 keV appears when the 3--100 keV best 
fit spectra are extrapolated to the LECS energy band (see Figure 
\ref{soft excess}).
With the cold reflection factor $R$, the energy and  width of 
the  line fixed at the values from the previous fit, we first
attempted to model the soft excess with one black body component:
the result is not good  
($(\chi^2/dof)_{2000}$=186.7/143, $(\chi^2/dof)_{1998}$=166.1/143);
a much better result in both observations is obtained with the 
addition of a second black body (the probability of exceeding F with 
the addition of two parameters is
greater than 99.9 per cent in both spectra). It is worth
recalling here that Pounds et al. (2001) resorted to three black bodies
to model the soft excess observed with \xmm.

When we try to substitute the two black body components with a disc black body model, the fit becomes worse in both spectra ($\chi^2/dof = 177/140 $ in 2000 and $\chi^2/dof = 150/140$ in 1998).

We also tried, without success, the addition, instead of the second black 
body, of two edges at the energies of OVII (0.74 keV) and 
OVIII (0.87 keV), whose optical depths turned out consistent with 
zero in both spectra.

\begin{figure*}
\centering
\includegraphics[height=8.cm,width=7cm]{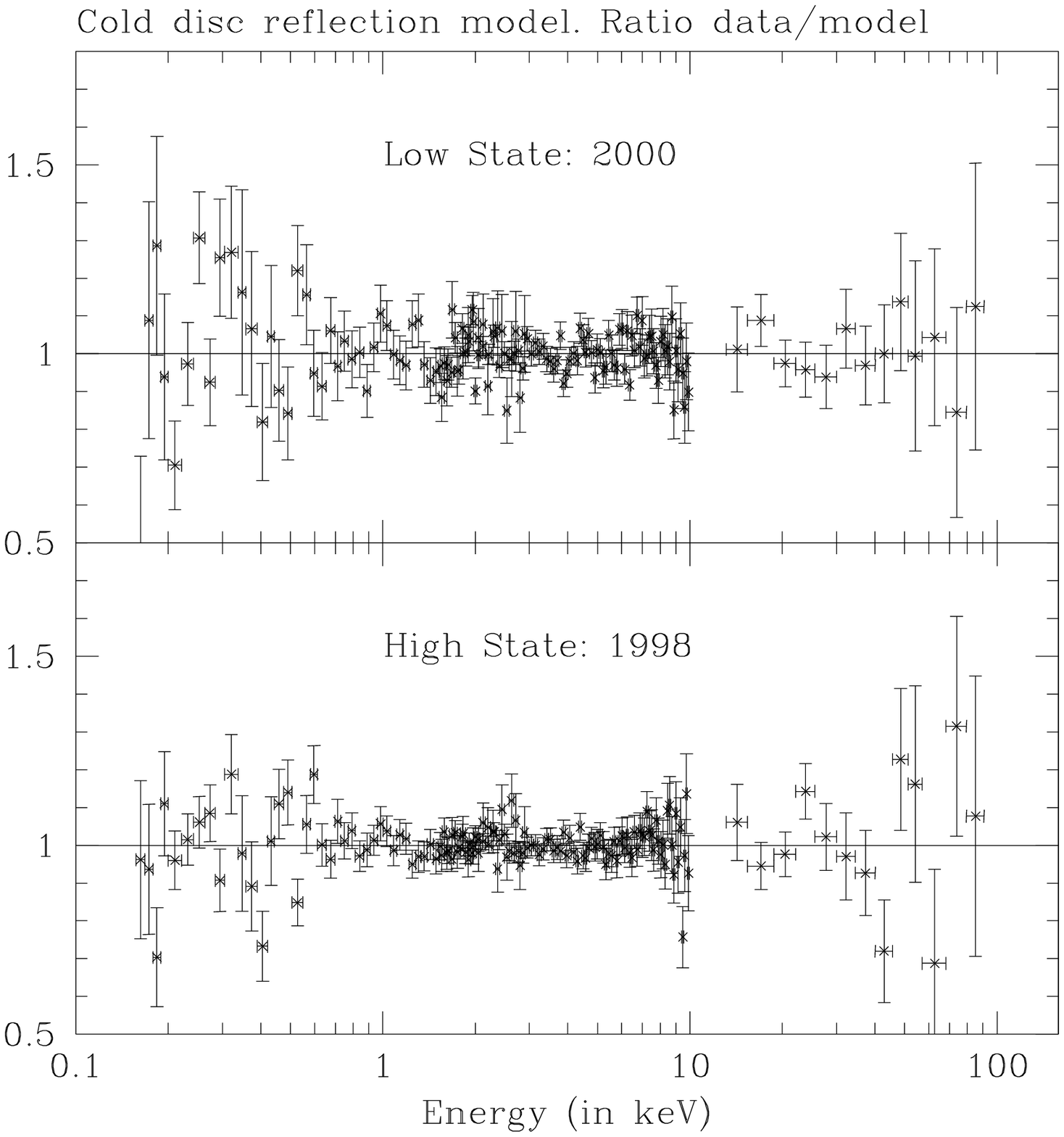}
\includegraphics[height=8.cm,width=7cm]{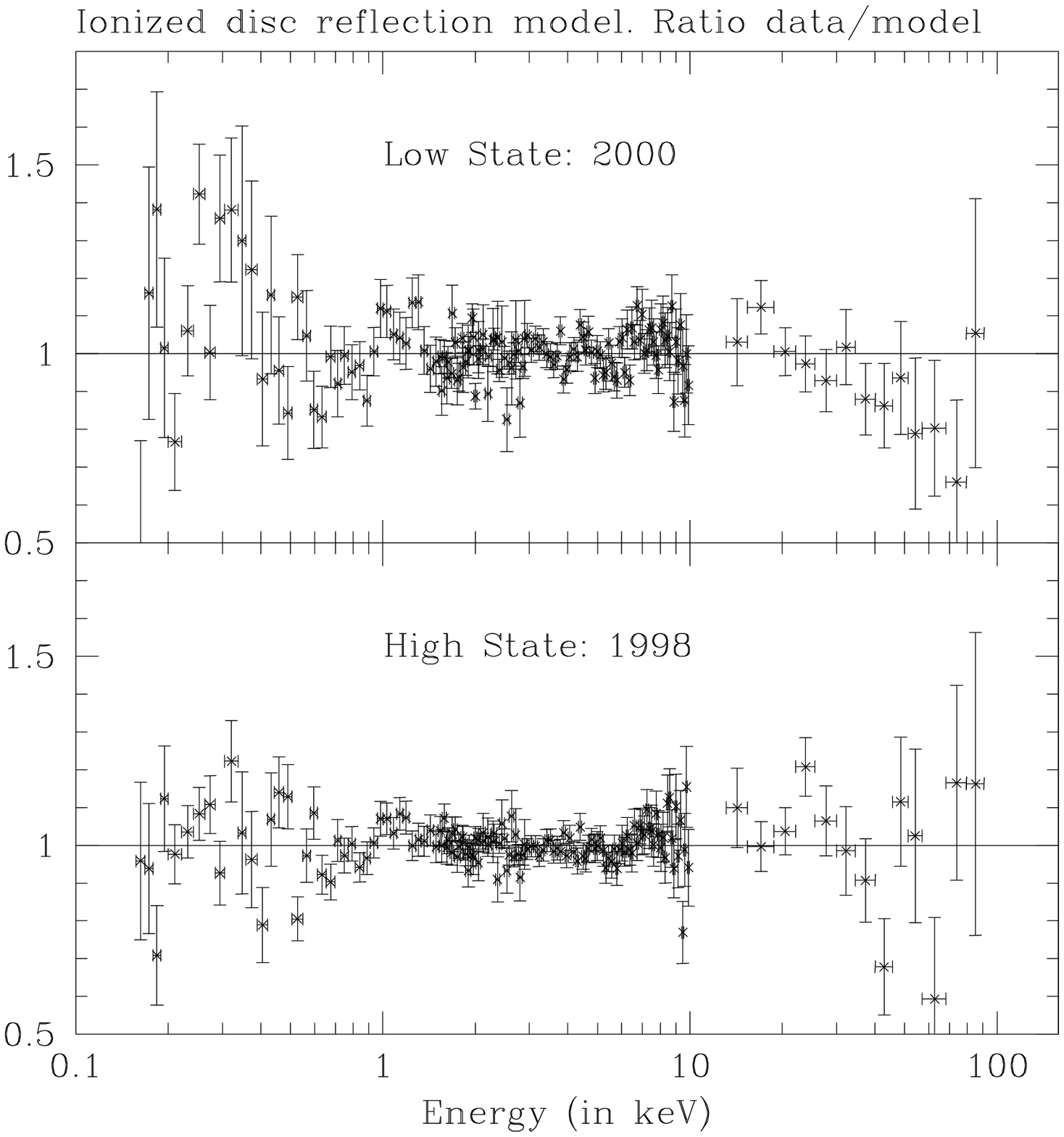}
\caption[]{Data/model ratio when  the spectra in the low (upper panels) 
and high (lower panels) flux 
states of \509 are reproduced with a cold (left) and ionized (right) 
disc reflection model. See Table \ref{cold disc model} and \ref{iondisc table} and for best fit parameters}
\label{iondisc_ratio}
\end{figure*}

The best fit spectra with two black bodies are summarized
in Table \ref{cold disc model}, and the data/model ratio is shown 
in the left panel of Figure \ref{iondisc_ratio}.
The intensities of the iron line in the two flux states are consistent 
with a constant value:  
I$_{Fe}^{2000}=(2.8 \pm 1.1) \times 10^{-5}$ ph cm$^{-2}$ s$^{-1}$ and
I$_{Fe}^{1998}=(3.4 \pm 2.0) \times 10^{-5}$ ph cm$^{-2}$ s$^{-1}$.

If, following the claim by Pounds et al. (2001), a second iron line is added
to the previous model to account for a 
possible contribution from an ionized component,
the fit improves marginally (the probability to exceed F with the addition 
of two parameters, the
energy and intensity of the line, is 95 and 93 per cent in the low and high
flux state respectively). In both spectra the (poorly constrained) line
energy is consistent with H-like iron, and the equivalent widths are 
$\sim$ 60 eV and $\sim$ 45 eV in the low and high flux states, respectively.

\begin{figure*}
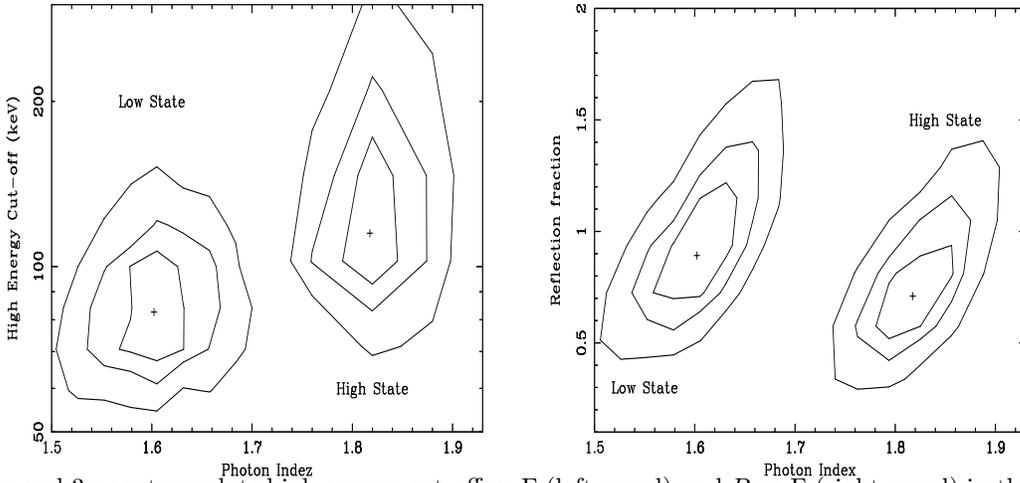

\centering
\includegraphics[height=6cm,width=6.cm,angle=-90]{ec_gamma_tot_pexrav.ps}
\hspace{1cm}
\includegraphics[height=6cm,width=6.cm,angle=-90]{r_gamma_tot_pexrav.ps}
\caption[]{1$\sigma$, 2$\sigma$ and 3$\sigma$ contour plots high energy cut-off
vs $\Gamma$ (left panel) and $R$ vs $\Gamma$ (right panel) in the two flux 
states when the spectra 
are fitted with a cold disc reflection model.}
\label{gamma_ec}
\end{figure*}

The results shown in Table \ref{cold disc model} quantitatively 
confirm our first comment on Figure 1, with a change of the spectral index 
$\Delta\Gamma \sim 0.2$ (see Figure \ref{gamma_ec}).
This behaviour is consistent with that observed in other
Seyfert 1 galaxies (NGC~4151: Perola \etal 1986, Piro \etal 1998; NGC~5548: Nicastro \etal 2000;
IC~4329a: Done \etal 2000; NGC~7469: Nandra \etal 2000; MCG~+6-30-15:
Vaughan \& Edelson 2001; NGC~3783: De Rosa \etal 2002a), 
and with the F$_x$ vs $\alpha$ relation
expected in a Comptonization model for the intrinsic continuum emission
(Petrucci \etal 2000 and references therein).

With regard to our second comment to Figure 1, we note
that the strength of the two black body soft excess, in the 0.25--2 keV range,
is a factor about three times larger in the 1998 epoch 
than in the 2000 epoch.

With regard to our third comment to Figure 1,
the normalization of the Compton reflection component
(which we verified to depend very weakly on the value of the slope
of the incident radiation) turns out to be 
A = $(2.3 \pm 0.6) \times 10^{-2}$ in 1998  and
A = $(2.0^{+0.4}_{-0.6}) \times 10^{-2}$ in 2000,
that is consistent with a constant value, very much like
the intensity of the cold iron line. 

The fact that, as shown in Figure 4, its relative normalization appears
also constant (that, within the large errors, could suggest 
a reflection component which follows the intrinsic changes), 
is only the consequence of the F$_x$ vs $\alpha$ relation holding 
in the flux range that we happened to cover.

\subsection{The ionized disc reflection model}
\label{ionized disc section}

The possible presence of a line from highly ionized iron 
(as claimed by Pounds \etal 2001) is suggestive of 
reflection from an ionized accretion disc. 
In this case the reflected component can be strong 
below $\sim$ 1 keV and then significantly contribute to the 
observed soft excess in \509.

We therefore fitted the spectra with
a reflection model that takes into account ionization
of the accretion disc
(Ross \& Fabian 1993). The most important quantity in determining
the shape of the reflected continuum is the ionization parameter
$\xi=4\pi F_x/n_H$,
where F$_x$ is the X-ray flux (between 0.01-100 keV) illuminating a slab of
gas with solar abundance and constant hydrogen number density
n$_H$=10$^{15}$ cm$^{-3}$. The incident flux is assumed to be a power law
with spectral photon index $\Gamma$ and a sharp high energy 
cut-off at 100 keV. The reflected spectrum, which includes the Fe K$\alpha$, 
is multiplied by
$R$ and added to the primary continuum.
Different values of $\xi$ indicate different
ionized states, affecting the strength and width of the Fe line
(Matt \etal 1993, Matt \etal 1996), and of the absorption edges.
%The soft emission from the accretion disc is not included in the models.
We applied to the reflection spectrum the relativistic smearing
for the Schwarzschild metric assuming an emissivity law 
r$^{\beta}$, with $\beta$=--3 and inner and outer radii fixed to 
6r$_g$ and 1000r$_g$. Finally, a narrow and cold iron
line was further added to the fitting model.

The best fit parameters are given in Table \ref{iondisc table},
and the data/model ratios are plotted in the right panel of 
Figure \ref{iondisc_ratio}. Notably, in the high flux level 
spectrum (1998) the ionized reflection accounts for the soft
X-ray emission, and there is no need for an additional soft
component. The $\chi^2$ is acceptable, albeit not
as good as with the alternative model; the same is not
true in the low flux level spectrum (2000), where in particular
the residuals both below 1 keV and above 20 keV show a
systematic deviation, and the $\chi^2$ is very poor.

The ionization parameter in both spectra is
$\sim$ 40 erg cm s$^{-1}$. For such a value, according to  
Matt \etal (1993), 
the disc is expected to contribute a neutral iron line of about 100 eV.
%EW$_{discline}^{2000}\sim$ 100 eV and EW$_{discline}^{1998}\sim$ 80 eV,
Indeed, we find that the intensity of the narrow line 
is somewhat smaller than
that measured in the cold disc model.  Again, the intensity of the narrow line 
does not change from one observation to the other, as expected if the
line is produced in distant matter.

To test the possibility that the distant matter responsible
for the narrow line were optically thick,
we added a constant cold reflection component.
In the high flux spectrum the reflection at low energies
does still account reasonably well for the soft excess and the fit remains 
acceptable ($\chi^2/dof$=158.5/142), although the cold component becomes
dominant over the ionized one  
($R_{cold}=1.1^{+0.65}_{-0.25}$, $R_{ion}=0.4\pm 0.2$),
while in the low flux spectrum no improvements is
attained at low energies and the fit is worse 
($\chi^2/dof$=187/143).

\begin{figure*}
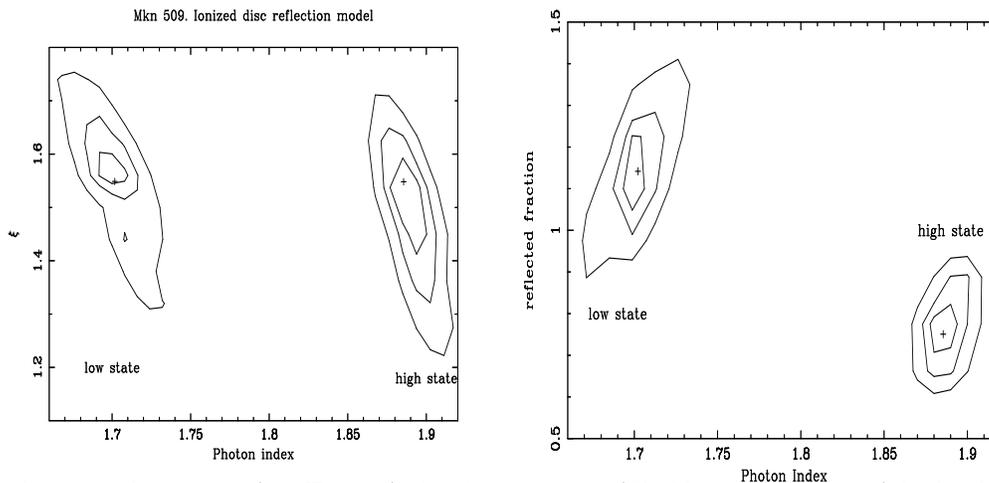

\centering
\includegraphics[height=6cm,width=6cm,angle=-90]{xi_gamma.ps}
\hspace{1cm}
\includegraphics[height=6cm,width=6cm,angle=-90]{refl_gamma_tot.ps}
\caption[]{Ionized disc reflection model (see Table \ref{iondisc table}). 
Confidence plots (68, 90 and 99 per cent) ionization parameter log $\xi$ 
vs Photon index $\Gamma$ (left panel) and Reflected fraction vs $\Gamma$
(right panel) in the different flux states of \509.}
\label{xi-r-gamma}
\end{figure*}

We note that also with this model the fit bears out
a change $\Delta\Gamma$ $\sim$ 0.2 (see Figure 
\ref{xi-r-gamma} and Table \ref{iondisc table}).

On balance, it seems to us that the cold reflection model,
with the addition of an independent soft-excess component,
is better suited to describe the two observations than
the model just described.

\section{Discussion}

\subsection{The intrinsic continuum}

The broad-band \sax spectra of \509 are dominated by a variable  
power law with an exponential cut--off. The $e$--folding energy is
found within the range of  values 
observed in other Sy 1s with \sax (Perola \etal 2002). 
This model is an approximation of a Comptonized spectrum.
In fact one of the emission mechanism which is believed to work in the 
central region of 
AGN is a two-phases model involving a hot corona emitting medium-hard 
X-rays by Comptonization and an optically thick layer 
which provides the soft photons to be Comptonized 
(Haardt \& Maraschi 1993). However, the heating mechanism of the
electrons, the geometry of the 
disc-corona configuration and the origin of the variability are still 
very uncertain.

There is strong evidence
that in Seyfert 1 galaxies the X-ray spectrum softens as the 2-10 keV flux 
increases (e.g. 1H~0419-577, Page \etal 2002; NGC~3783, De Rosa \etal 2002a, 
MCG-6-30-15,  Vaughan \& Edelson 2001;
NGC~5548, Petrucci \etal 2000, Nicastro \etal 2000; NGC~7469, 
Nandra \etal 2000; IC~4329A, Done \etal 2000, NGC~4151 Perola \etal 1986).
Our analysis confirms this behaviour also in the case of \509, where a flux 
variation by a factor two in the 2--10 keV range is observed along
with a change of the power law slope  $\Delta\Gamma\sim$0.2 
(see Tables \ref{cold disc model}, 
\ref{iondisc table} and Figures \ref{gamma_ec}, \ref{xi-r-gamma}).

Analysing the spectral variability with a Comptonization model 
(Haardt \& Maraschi 1993) we found that an additional black body component 
was required (with $P_F$ greater that 99.9 per cent), to model the soft X-ray 
spectra  in both observations. 
The fit is still good in 1998 
($\chi^2/dof = 130/141$), while this is not true in 2000 
($\chi^2/dof = 174/141$), where large deviations are present below 
1 keV and above 40 keV.
We found that for the high state  the temperature of the corona 
and the Comptonization parameter $y$  (the Comptonized to soft luminosity 
ratio), are lower than in the low state:
$kT_e=65\pm 10$ keV, and $\tau=1.00^{+0.08}_{-0.06}$ in 1998 and $kT_e=92^{+20}_{-10}$ keV and $\tau=0.88^{+0.06}_{-0.04}$ in 2000.
This behaviour, together with the fact that the continuum variation is higher 
at low energies (see lightcurves in Figure \ref{lightcurve}), suggests an 
increase, in the high state, of the seed photons able to Comptonize the corona.
This increase could be due 
either to some geometrical effect (e.g. a change of the inner radius 
of the disc) or to a change of the albedo of the 
disc (e.g. a variation of the ionization state). 
In both the hypotheses some variations in the strength of the reflection 
features are expected. However our analysis is consistent 
with a constant value of the reflection components (see discussion in 
Section \ref{refl discussion}). Then our analysis supports the idea that the 
observed variability is driven by a change of the soft photon flux 
which is not due to a change of either the geometry or the albedo of the disc.

\subsection{The soft X-ray excess}
\label{se discussion}

As indicated by \xmm and  \sax
observations, the soft excess is a common component in Seyfert 1
spectra (see e.g. Pounds \& Reeves 2002).  Whether this
component represents thermal emission directly from the
accretion disc, or the result of the disc reprocessing of the hard 
X-rays impinging upon it, is still unclear.
In a study of a small sample of Sy 1s observed by \xmm,
Pounds \& Reeves (2001) conclude that the ``gradual soft excess''
(GSX) detected in \509 could be interpreted as Comptonized thermal
disc emission, this view being supported by the absence of discrete
spectral features in RGS data.  In addition they suggest that this
broad soft emission component might be a common feature, which can sometimes
be obscured by a warm absorber.

The analysis of the two \sax spectra of \509 shows that the absolute
strength of the soft excess is definitely higher in the high flux state. 
However when it is reproduced with two black body components,
it accounts for $\sim$ 30 per cent of the 0.25-2 keV observed luminosity 
in 2000 ($L_{0.25-2 keV}^{2000}= 9\times 10^{43}$ erg s$^{-1}$) and 
less than 25 per cent in 1998 ($L_{0.25-2 keV}^{1998}= 2.4\times 10^{44}$ 
erg s$^{-1}$).
When the spectra are fitted with an ionized disc reflection
model, the soft excess in the high flux level state (1998)
can be completely accounted for by the
reflection of the disc, with no additional components required. 
In the low flux level state (2000), instead,
the stronger relative contribution of the soft excess is 
only partially accounted for 
by a larger value of the reflection fraction (from 
a disc surprisingly characterized by the same ionization parameter as in 1998),
and the global fit can hardly be considered acceptable.
Despite the attractive behaviour of this model in one of
the two observations, the alternative model with a
cold reflection and the soft additional component,
at least from a statistical point of view, behaves
well in both observations, and it offers a natural
explanation for the presence of the narrow and cold iron line.

We furthermore note that 
the black body components do not represent the unique empirical 
model which provides a good fit of the soft
excess.  Perola \etal (2000) describe this component with a soft
power law, and find that its extrapolation to lower energies 
matches the observed strength of 
the UV emission, admittedly observed in a different epoch. This
lends some support to the idea that it could arise from Comptonization in
the innermost part of the accretion disc (see also the case of NGC~5548 in
Pounds \etal 2002).

\begin{figure*}
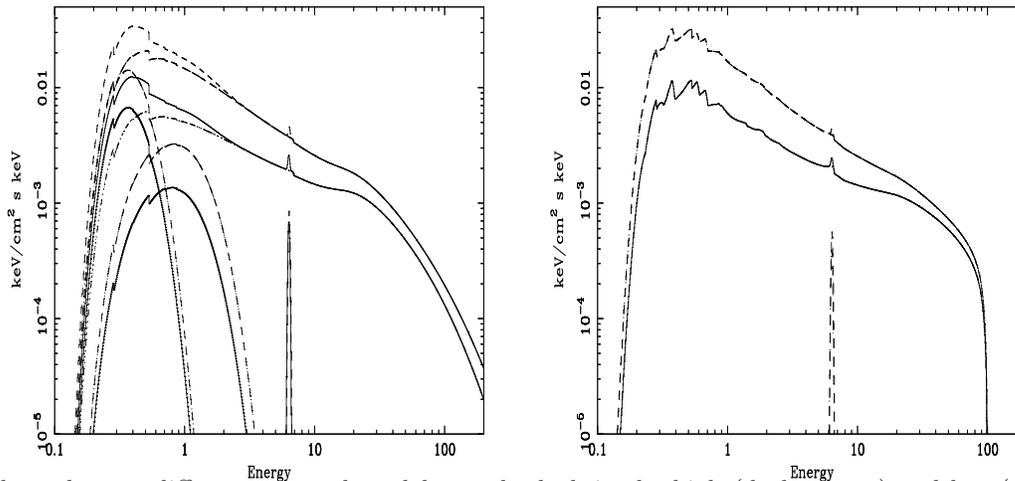

\centering
\includegraphics[height=6cm,width=6cm,angle=-90]{cold4_model.ps}
\hspace{1cm}
\includegraphics[height=6cm,width=6cm,angle=-90]{ion1_model.ps}
\caption[]{We show the two different spectral models we checked,
in the high (dash curves) and low (solid curves) flux state. 
In the left panel the cold disc reflection model is shown
together with the black body components to reproduce the soft X-ray spectra.
In the ionization disc reflection model (right panel) no additional component 
are required to model the spectra.}
\label{models}
\end{figure*}

\subsection{The reflection and fluorescent components}
\label{refl discussion}

Both the Compton reflection hump and the neutral iron emission line are
detected in the \sax observations of \509.  A single Gaussian component
reproduces well the line profile. Perola et al. (2000) got similar
results on the reflection features using a power law rather than
two black body components to model the soft excess.  The energy of the
iron line they found is slightly higher than the value we 
report here, because in our broad--band fit we kept the intrinsic 
width of the line fixed to 0.1 keV.

The energy of the iron line indicates a low stage of ionization (FeI-XVII). 
Its intensity does not vary significantly 
between 1998 and 2000 and, moreover, it is 
consistent with that measured in a simultaneous  
\chandra HEG and \rxte PCA observation (Yaqoob \& Padmanabhan 2003)  
when the flux was similar to that observed by \sax in 1998, 
F$_{2-10keV}^{Chandra}=5.5 \times 10^{-11} \flux$, and in a 
\xmm observation in 2000 (Pounds \etal 2001). The amplitude of the reflection
component is also consistent with a constant value in
the two \sax observations, despite the variations in the primary flux. 
Thus it is most likely that
both fluorescence and reflection take place at a distance of at least
about one parsec from the central source.
 
A narrow component of the iron line is a common feature in 
Seyfert 1s (Padmanabhan \& Yaqoob 2003 and references therein).
A recent analysis of a sample of nine Seyfert 1s (\509 included) 
observed by \xmm (Reeves 2003) found that the majority of the sources 
show an unresolved line at 6.4 keV. In general terms, given the typical
EW found, this line is not necessarily
produced by Compton thick material, for instance it could
come from the Broad Line Region 
(which have typical column densities of $10^{23} cm^{-2}$, Netzer 1990). 

The case of \509, as
assessed in this paper, adds to those of NGC~4051 (Guainazzi \etal 1998) 
and NGC~5506 (Matt \etal 2001), where a reflection component associated with
a narrow cold iron line has indeed been detected.

Concerning the very marginal evidence we found of a ionized
line in addition, it is worth mentioning the case of other
objects where such, and generally weak, line has been observed,
like NGC~5506 (Matt \etal 2001), NGC~7213 (Bianchi et al. 2003), Mkn~205 
(Reeves \etal 2001), NGC~3783 (De Rosa \etal 2002a), NGC~7469 
(De Rosa \etal 2002b). As already mentioned, in the first
\xmm observation of \509 in 2000 Pounds \etal (2001) claimed
the discovery of a weak ionized line in addition to the narrow cold
line. After a reanalysis of the same observation, along with the
analysis of a second one made in 2001, Page \etal (2003) reach
the conclusion that 
illumination of a distant, cold and optically thick material is a simpler 
and self consistent explanation of the iron spectral features, rather 
than models including reflection from an ionized relativistic disc.

\section{Conclusions}
Taking advantage of two 0.1--100 keV \sax observations of \509, performed
about two years apart, which found the source a factor of two
different in the 2-10 keV flux level, we could profitably 
compare two alternative models, a classical one with a cold
Compton reflection, already investigated on the first of the two
by Perola \etal (2000), the other with the reflection from
an ionized disc.

Common results of the two fitting models are:

a) The slope of the power law increases, $\Delta\Gamma\sim$0.2 ,
when the 2--10 keV flux is a factor two higher. This result
agrees with the behaviour found in several other objects
of this class, mentioned in the previous Section, and
lends another case to support the Comptonization models
for the origin of the power law, in the form applied e. g. to
NGC~5548 by Petrucci \etal (2000).

b) There is a cold iron fluorescent line, whose
intensity and width are consistent with results
obtained on the same object with \xmm (Pounds \etal 2001;
Page \etal 2003) and with a simultaneous  
\chandra HEG and \rxte PCA observation (Yaqoob \& Padmanabhan 2003).
The stability in the intensity of this line, despite the rather
large variations of the continuum level, speaks definitely
in favour of a distant (order one parsec) placement
of the fluorescent material.

With respect to a simple power law, at low energies and in
both observations there is a fairly strong and rather broad
soft excess. The ionized disc reflection model can in 
principle account self-consistently for the existence of
such an excess. And indeed, in the high state observation
(1998) this model gives a fairly good fit, without
requiring any extra component, except for the narrow line
mentioned above. However, in the low state observation (2000)
the outcome is a very poor fit, with substantial positive
residuals below 1 keV. From a purely empirical point of view,
this difference between the two states follows from the
relative contribution of the soft excess being stronger when the source was
in the low state, when the power law was at the same time
harder.

With the cold reflection model, the soft excess needs necessarily
be modelled with an additional component, for which an
empirical combination of two black bodies gives fair results
in both states (an equally fair result was obtained on the
first observation by Perola \etal, 2000, adopting a soft
power law). The luminosity of this excess in the 0.25--2 keV range
turns out to be about three times higher in the high (1998)
state than in the low (2000) state. In our mind its origin
and behaviour remains a substantially obscure issue.

The EW of the narrow line is modest
and, in principle, could be explained as fluorescence from Compton
thin material, for instance in the Broad Line Region.
However a very important result with the cold reflection model is that
the normalization of the Compton hump is perfectly
consistent (within the large errors),
with a constant value between the two epochs.
This further argues in favour of a common 
origin of both the line and the reflection component in a 
distant, Compton--thick material. 
The simultaneous measurement of both the line and the
hump, which could be safely afforded with
\sax and now with \rxte (alone or in a simultaneous combination with
\xmm and \chandra), adds \509 to the
still short list of cases (NGC~5506, Matt \etal 2001;
NGC~4051, Guainazzi \etal 1998) in which the above
conclusion could be credibly reached.

%\begin{acknowledgements}
%We gratefully acknowledge A.C. Fabian and D.R. Ballantyne for 
%providing us with the ionized disc reflection code. 
%A.D.R. would like to thank P. Grandi for useful discussions. 
%We thank the SAX Scientific Data Center and
%the anonymous referee who provided helpful suggestions and comments.
%\end{acknowledgements}

\end{document}